\documentclass[twocolumn,aps,prl,showpacs]{revtex4}

\usepackage{graphicx}% Include figure files
\usepackage{dcolumn}% Align table columns on decimal point
\usepackage{bm}% bold math
\usepackage[latin1]{inputenc}

\def\be{\begin{equation}}
\def\ee{\end{equation}}
\def\bea{\begin{eqnarray}}
\def\eea{\end{eqnarray}}

\begin{document}

\title{$N$-particle $N$-level Singlet States:
Some Properties and Applications}
\author{Ad\'{a}n Cabello}
\email{adan@us.es}
\affiliation{Departamento de F\'{\i}sica Aplicada II,
Universidad de Sevilla,
41012 Sevilla, Spain}
\date{\today}
%First version: August 15, 2001.
%This version: August 15, 2002.
%After PRL's proofs.
%Phys. Rev. Lett. {\bf 89}, 100402 (2002).
%This will be quant-ph/0203119v4.

%%%%%%%%%%%%%%%%%%%%%%%%%%%%%%%%%%%%%%%%%%%%%%%%%%%%%%%%%%%%%%%%%%%%%

\begin{abstract}
Three apparently unrelated problems which have no solution using classical tools
are described: the ``$N$-strangers,'' ``secret sharing,''
and ``liar detection'' problems. A solution for each of them
is proposed. Common to all three solutions is the use
of quantum states of total spin zero of $N$ spin-$(N-1)/2$ particles.
\end{abstract}

%Quantum cryptography
%03.65.Ta}
%Foundations of quantum mechanics; measurement theory
\maketitle

%%%%%%%%%%%%%%%%%%%%%%%% Introduction %%%%%%%%%%%%%%%%%%%%%%%%

Not long ago, during a meeting on quantum information, a speaker
asked the participants to make a list of ``interesting'' quantum
states, namely, those which have potential applications
(particularly tasks which were impossible using classical physics)
or illustrate fundamental issues of quantum mechanics
\cite{Cirac00}. The final list was rather short:
Einstein-Podolsky-Rosen-Bohm-Bell states of two particles
\cite{EPR35,Bohm51,Bell64}, Greenberger-Horne-Zeilinger (GHZ)
states of three or more qubits \cite{GHZ89}, Werner (mixed) states
\cite{Werner89}, Hardy states (pure nonmaximally entangled states
of two particles) \cite{Hardy}, Horodecki ``bound'' states
(entangled mixed states from which no pure entanglement can be
distilled) \cite{HHH98}, and $W$ states of three qubits
\cite{DVC00}. Curiously, most of them were first introduced in
connection with Bell's theorem, and some appeared in the course of
the classification of entanglement. Surprisingly, none of them
was originally introduced as the answer to a practical problem
without classical solution (although most of them have later found
numerous applications
\cite{Ekert91,BW92,BBCJPW93,CB97,ZZHW98,HBB99,KKI99,LK00,Cabello00,XLG01}).

Here we shall introduce three apparently unrelated problems without
classical solution and then propose a solution
for all of them, using quantum mechanics.
Common to all these solutions is the use of a
family of quantum states.

%%%%%%%%%%%%%%%%%%%%%%%%%%%%%%%%%%%%%%%%%%%%%%%%%%%%%%%%%%%%%%%

{\em (1) The $N$ strangers problem.}---The scenario for this problem
is an extension to a high number $N$ of players of the situation
described in Patricia Highsmith's novel and Alfred Hitchcock's
movie {\em Strangers on a Train} \cite{HH50}: $N$ complete
strangers $A_i$ ($i=1,\ldots,N$), meet on a train. $A_i$ wants
$B_i$ dead. During small talk, one suggests that an ``exchange''
murder between $N$ complete strangers would be unsolvable.
After all, how could the police find the murderer when he/she is a total and complete
stranger with absolutely no connection whatsoever to the victim?
$A_i$ could kill $B_k$, etc. \cite{commmurder}.
However, such a plan suffers from an important problem:
if all the players know who the murderer of each victim is,
then the whole plan is vulnerable to individual denunciations.
Alternatively,
if the distribution of victims is the result of a secret lottery, how
could the murderers be assured that the lottery was not rigged and that nobody had
contrived the result or could ascertain it?

The problem is then how to distribute the victims $\{B_i\}_1^{N}$
among the murderers $\{A_i\}_1^{N}$, which share no previous
secret information nor any secure classical channel, in a way that
guarantees that each murderer $A_i$ knows only the identity of
his/her victim and that nobody else (besides the murderers) knows
anything about the assignment of the victims.

%%%%%%%%%%%%%%%%%%%%%%%%%%%%%%%%%%%%%%%%%%%%%%%%%%%%%%%%%%%%%%%

{\em (2) The secret sharing problem.}---This problem was already described,
for $N=3$, in \cite{HBB99}.
It could arise in the following context:
$A_1$ wants to have a secret action taken on her behalf at a distant
location. There she has $N-1$ agents, $A_2$, $A_3$, \ldots, $A_N$ who carry it out
for her. $A_1$ knows that some of them are dishonest,
but she does not know which one it is. She cannot simply send a secure
message to all of them, because the dishonest ones will try to
sabotage the action, but
it is assumed (as in \cite{HBB99}) that if all of them carry it
out together, the honest ones will keep the
dishonest ones from doing any damage.

The problem is then that $A_1$ wishes to convey a cryptographic
key to $A_2$, $A_3$, \ldots, $A_N$ in such a way that none of them
can read it on their own, only if all the $A_i$ ($i=2,3,\ldots,N$)
collaborate. In addition, they wish to prevent any eavesdropper
from acquiring information without being detected. It is assumed
that $A_1$ shares no previous secret information nor any secure
classical channel with her agents.

Different quantum solutions to this problem for $N=3$
has been proposed using
either GHZ \cite{ZZHW98,HBB99} or Bell states \cite{KKI99}.
Below we shall propose a different solution for any $N$
which exhibits some additional advantages.

%%%%%%%%%%%%%%%%%%%%%%%%%%%%%%%%%%%%%%%%%%%%%%%%%%%%%%%%%%%%%%%

\begin{figure}
\centerline{\includegraphics[width=8.2cm]{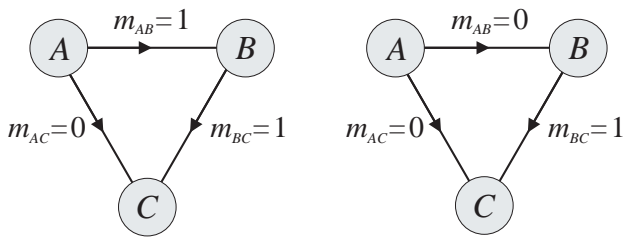}}
\caption{\label{Liar}
The liar detection problem.
Left: $A$ is a liar because she sends different messages to $B$ and $C$.
Right: $B$ is a liar because he sends $C$ a message different
to the one he received from $A$.
The task is for $C$ to identify who is being dishonest, $A$ or $B$.}
\end{figure}

{\em (3) The liar detection problem.}---Let us consider the
following scenario: three parties $A$, $B$, and $C$ are connected
by secure pairwise classical channels. Let us suppose that $A$
sends a message $m$ to $B$ and $C$, and $B$ sends the same message
to $C$. If both $A$ and $B$ are honest, then $C$ should receive
the same $m$ from $A$ and $B$. However, $A$ could be dishonest and
send different messages to $B$ and $C$, $m_{AB} \neq m_{AC}$
(Fig.~\ref{Liar}, left), or, alternatively, $B$ could be dishonest
and send a message which is different to the one he receives,
$m_{BC} \neq m_{AB}$ (Fig.~\ref{Liar}, right). For $C$ the problem
is to ascertain without any doubt who is being dishonest. This
problem is interesting for classical information distribution in
pairwise connected networks. The message could be a database and
the dishonest behavior a consequence of an error during the
copying or distribution process. This problem has no solution by
classical means. It is at the heart of a slightly more complicated
problem in distributed computing called the Byzantine agreement
problem \cite{LSP82}, a version of which has been recently solved
using quantum means by Fitzi, Gisin, and Maurer \cite{FGM02}.
Indeed, the solution for our liar detection problem is based on
theirs.

%%%%%%%%%%%%%%%%%%%%%%%%%% Figure %%%%%%%%%%%%%%%%%%%%%%%%%%

%%%%%%%%%%%%%%%%%%%%%%%%%%%%%%%%%%%%%%%%%%%%%%%%%%%%%%%%%%%%%%%

The next step is to show that all of these problems can be
solved if each of the $N$ participants are in possession of
a sequence of numbers with the following properties:
(i) It is {\em random} (i.e.,
generated by an intrinsically unrepeatable method which
gives each possible number with the same probability of
occurrence).
(ii) The possible numbers are integers from $0$ to $N-1$.
(iii) If a number $i$ is at
position $j$ of the sequence of party $k$, $i$ is not at position $j$ in
the sequence of a different party.
(iv) Each party knows only his/her own sequence.
(v) Nobody else (besides the parties) knows
the sequences.
Properties (iv) and (v) are difficult to accomplish using
classical tools
due to the fact that information transmitted in classical form can be
examined and copied without altering it in any detectable way.
However, as quantum key distribution protocols show \cite{BB84,Ekert91},
quantum information does not suffer from such a drawback.

Assuming we have a reliable method to generate
sequences of numbers
with properties (i) to (v) among $N$ distant parties,
a method that will be presented below,
then the solutions to the above problems are as follows.

%%%%%%%%%%%%%%%%%%%%%%%%% Solution to 1 %%%%%%%%%%%%%%%%%%%%%%%%%

{\em Solution to the $N$ strangers problem.}---Each victim $B_i$
is assigned a label, taken from $0$ to $N-1$.
If murderer $A_i$'s sequence starts with $j$, then
$A_i$ must kill $B_j$, etc.
The remaining entries of the sequence can be
used for subsequent rounds.
The result tells every murderer
who his/her victim is in such a way that prevents any
murderer (or even a small group of them) from denouncing
or blackmailing another.
The only way to ascertain with certainty who murdered $B_j$ is
that all the other murderers confess \cite{commown}.

%%%%%%%%%%%%%%%%%%%%%%%%% Solution to 2 %%%%%%%%%%%%%%%%%%%%%%%%%

{\em Solution to the secret sharing problem.}---The key is
defined as $A_1$'s sequence.
The only way to reveal it is to make the remaining $N-1$ parties
share their respective sequences;
the key is then composed by the missing results.
If a
dishonest party $D$ declares a result which is different to his/her
actual result, then there is a probability $1/(r-1)$, where $r$ is the
number of honest parties which have not yet declared their
results, that other honest party $H$ has obtained that result. Then
$H$ would stop the process, so Alice's key (and thus Alice's action)
would remain safe (dishonest parties cannot sabotage Alice's
action if they do not know what it is). The order in which the
agents declare their respective results must change from round to
round to avoid any dishonest party being always the last to
declare.

%%%%%%%%%%%%%%%%%%%%%%%%% Solution to 3 %%%%%%%%%%%%%%%%%%%%%%%%%

{\em Solution to the liar detection problem.}---Let us suppose that
the message $m$ is a trit value $0$, $1$, or $2$. The three
parties agree to use the following protocol:
(I) If the transmitted
message is $m_{ij}$, then the sender $i$ must also send $j$ the list
$l_{ij}^{(m_{ij})}$ of positions in his/her sequence in which the
number $m_{ij}$ appears. Note that
if the sequences are random and long enough
then any $l_{ij}^{(m_{ij})}$ must contain about one third
of the total length $L$ of the sequences.
(II) The receiver $j$ would not
accept any message if the intersection between the received list
$l_{ij}^{(m_{ij})}$ and his/her list $l_{j}^{(m_{ij})}$ is not
null nor if $l_{ij}^{(m_{ij})} \ll L/3$ elements.
We will assume that requirements (I) and (II) force the dishonest one to send
correct but perhaps incomplete lists.
Otherwise, if $i$ sends a list containing $n$ erroneous data, then
the probability that $j$ does not accept the message $m_{ij}$ would
be $(2^n-1)/2^n$.
In addition, (III) $B$ must
send $C$ the list $l_{BC}^{(m_{AB})}$ containing the sequence he
has (supposedly) received from $A$. Therefore, when $C$ finds that
$m_{AC} \neq m_{BC}$, she has received three lists to help her to
find out whether it is $A$ or $B$ who is being dishonest.

According to rules (I) to (III),
if $B$ wants to be dishonest $l_{BC}^{(m_{BC})} \cup
l_{BC}^{(m_{AB})}$ must necessarily be a subset of
$l_{B}^{(m_{BC})}$, because $B$ does not know $l_A^{(m_{BC})}$.
However, the length of $l_{B}^{(m_{BC})}$ is about $L/3$,
while $C$ is expecting $B$ to send her two lists with a total length of $2 L/3$; then $C$
would conclude that $B$ was being dishonest.
Alternatively, if it
is $A$ who is being dishonest, the lengths of the two lists that $C$ received from
$B$ would total about $2 L/3$; then $C$ would conclude that $A$
was being dishonest.

%%%%%%%%%%%% $N$-particle $N$-level singlet states %%%%%%%%%%%%

The next step is to present a method to generate
among $N$ distant parties
sequences of numbers
with properties (i) to (iv).
A possible quantum solution, probably not the only one,
but maybe the most natural, is by distributing among
all $N$ parties an
$N$-particle $N$-level singlet state of total spin zero.
For arbitrary $N$ these states can be expressed as
\be
\left|{\cal S}_N \right\rangle={1 \over \sqrt {N!}}
\sum_{\scriptscriptstyle{
{\stackrel{\scriptscriptstyle{\rm permutations}}
{{\rm of}\;01\ldots(N-1)}}}}
\!\!\!\!\!\!
(-1)^t\left| ij\ldots n \right\rangle
\label{psin}
\ee
where $t$ is the number of transpositions of pairs of elements
that must be composed to place the elements in
canonical order ({\em i.e.}, $0,1,2,\ldots,N-1$) and
$\left| 01\ldots (N-1) \right\rangle$ denotes the tensor
product state $\left| 0 \right\rangle \otimes \left| 1
\right\rangle \otimes \cdots \otimes \left| N-1 \right\rangle$.
Particular examples of $|{\cal S}_N\rangle$ are as follows:
\begin{eqnarray}
\left|{\cal S}_2 \right\rangle & = & {\scriptstyle{1 \over \sqrt {2}}} \left(
\left| 01 \right\rangle - \left| 10 \right\rangle
\right), \label{psi2} \\
%%%%%%%%%%%%%%%%%%%%%%%%%%%%%%%%%%%%%%%%%%%%%%%%%%%%%%%%%%%%%%%
\left|{\cal S}_3 \right\rangle & = & {\scriptstyle{1 \over \sqrt {6}}} ( \left|
012 \right\rangle - \left| 021 \right\rangle - \left| 102
\right\rangle + \left| 120 \right\rangle + \nonumber \\ & & \left|
201 \right\rangle - \left| 210 \right\rangle ), \label{psi3} \\
%%%%%%%%%%%%%%%%%%%%%%%%%%%%%%%%%%%%%%%%%%%%%%%%%%%%%%%%%%%%%%%
\left|{\cal S}_4 \right\rangle & = & {\scriptstyle{1 \over \sqrt {24}}} ( \left|
0123 \right\rangle - \left| 0132 \right\rangle - \left| 0213
\right\rangle + \left| 0231 \right\rangle + \nonumber \\ & &
\left| 0312 \right\rangle - \left| 0321 \right\rangle
%%%%%%%%%%%%%%%%%%%%%%%%%%%%%%%%%%%%%%%%%%%%%%%%%%%%%%%%%%%%%%%
-\cdots + \left| 3210 \right\rangle ). \label{psi4}
\end{eqnarray}
If we identify subsystems with spin-$(N-1)/2$ particles and
associate the state $\left| s \right\rangle$ with the eigenvalue
$s-(N-1)/2$ of the spin observable in some fixed direction, then
$|{\cal S}_N\rangle$ is the only state of $N$ particles of spin-$(N-1)/2$ which has
total spin zero.

The described
solutions assume that the $N$ parties share a large collection of
$N$-level systems in the $|{\cal S}_N\rangle$ state.
This requires a protocol to distribute and test
these states between the $N$ parties such that at the end of the
protocol either all parties agree that they share a $|{\cal S}_N\rangle$ state (and
then they can reliably apply the described solutions), or all of
them conclude that something went wrong (and then abort any
subsequent action). For $N=3$ such a distribute-and-test protocol is
explicitly described in \cite{FGM02} and can be easily be generalized
to any $N>3$.
The test requires that the parties compare a sufficiently large subset of their
particles which are subsequently discarded.

%%%%%%%%%%%%%%%%%%%%%%%% Correlations %%%%%%%%%%%%%%%%%%%%%%%%

For our purposes,
some interesting properties of the states $|{\cal S}_N\rangle$ are:

(a) They provide {\em correlated} results.
As can be easily seen in Eq.\ (\ref{psin}),
whenever the $N$ parties measure
the spin of the $N$ separated particles
along the direction used in Eq.\ (\ref{psin}), each of them
finds a different result in the set $\{0,\ldots,N-1\}$; thus such
results satisfy requirements (ii) and (iii).

%%%%%%%%%%%%%%% N-lateral rotational invariance %%%%%%%%%%%%%%%

(b) Moreover, $|{\cal S}_N\rangle$ are {\em $N$-lateral rotationally invariant}.
This means that if we act on any of them with the tensor product of
the $N$ rotation operators referring to all the particles for any
arbitrary rotation, the result will be to reproduce the same state
(within a possible phase factor).
Therefore, whenever the $N$ parties measure
the spin of the $N$ separated particles
{\em along any direction}
(but it must be the same direction for everyone), each of them
finds a different result in the set $\{0,\ldots,N-1\}$; thus such
results satisfy
requirements (ii) and (iii).
Therefore,
the direction of measurement could be randomly chosen and
publicly announced (once the particles have been distributed
among the parties) before any set of measurements.

%%%%%%%%%%%%%%%%%%%%%%% Nonseparability %%%%%%%%%%%%%%%%%%%%%%%

(c) In order to accomplish (i), (iv), and (v), an
essential property is {\em nonseparability}, that is, the quantum
predictions for the states $|{\cal S}_N\rangle$ cannot be
reproduced by any local hidden variables model in which the
results of the spin measurements are somehow determined before the
measurement. To show the nonseparability of
$|{\cal S}_N\rangle$ we have to study whether they violate Bell's
inequalities derived from the assumptions of local realism. Most
Bell's inequalities require two
alternative local {\em dichotomic} (taking values $-1$ or $1$) observables
$A_j$ and $B_j$ on each particle $j$.
To test nonseparability, we will use the dichotomic local observables
proposed by Peres in \cite{Peres92}. A Peres' observable $A_k$ can be
operationally defined as follows: to measure $A_k$, first measure
the spin component of particle $k$ along direction $A$, $S_A^{(k)}$.
If particle $k$ is a spin-$s$ particle, then measuring $S_A^{(k)}$
could give $2 s+1$ different results. Then assign value $1$ to
results $s$, $s-2$, etc., and value $-1$ to results
$s-1$, $s-3$, etc. The operator representing observable $A_k$
can be written as
\be \hat A_k = \sum_{m=-s}^{s} (-1)^{s-m} |S_A^{(k)}=m
\rangle \langle S_A^{(k)}=m |, \label{Peresob}
\ee
where
$|S_A^{(k)}=m\rangle$ is the eigenstate of the spin component along
direction $A$ of particle $k$.

Probably the simplest way to show the nonseparability \cite{comnonsep}
of the $|{\cal S}_N\rangle$ states is by considering the following scenario:
$N$ distant observers share $N$ $N$-level particles
in the $|{\cal S}_N\rangle$ state;
the $N-m$
%($1<m \le N-m$)
observers can
choose between measuring $A_j=A$ and $B_j=a$; the remaining $m$ observers can
choose between measuring $A_k=B$ and $B_k=b$.
Then nonseparability can be tested by means
of the following Bell's inequality, which generalizes to $N$ particles
the Clauser-Horne-Shimony-Holt (CHSH) inequality \cite{CHSH69}
\bea
|E_N(A,\ldots,A,B, \ldots,B)+ E_N(A,\ldots,A,b,\ldots,b)+ \nonumber \\
E_N(a,\ldots,a,B,\ldots,B)-E_N(a,\ldots,a,b,\ldots,b)| \le 2.
\label{CHSHg}
\eea
Note that this inequality uses only a subset of all possible
correlation functions (for instance, it does not use $E_N(A,a,\ldots,a,B,\ldots,B)$).
Restricting our attention to Peres' observables,
for the states $|{\cal S}_N\rangle$,
the correlation function $E_N^{(m)}(A,\ldots,A,B,\ldots,B)$,
which represents the expectation value of this product of the results
of measuring, for instance, $N-m$ observables $A$, and $m$ observables
$B$ is given by
\bea
\!\!E_N^{(N-1)} &\!\! = &\!\! (-1)^{f(N/2)} {1 \over N}
{\sin (N \theta_{AB}) \over \sin \theta_{AB}}, \\
%E_N^{(2)} & = &
E_N^{(N-2)} &\!\! = &\!\! (-1)^{f(N/2)} {1 \over N+2}
\left\{1+{\sin [(N+1) \theta_{AB}] \over \sin \theta_{AB}}\right\},
\eea
where $\theta_{AB}$ is the angle between directions $A$ and $B$
and $f(x)$ gives the greatest integer less than or equal to $x$.
In case of $m=1$, that is, using correlation functions of
the $E_N^{(N-1)}$ type, we have found that
states $|{\cal S}_n\rangle$ violate inequality (\ref{CHSHg})
for any $N$.
The maximum violation for $N=2$ is $2 \sqrt{2}$,
for $N=3$ is $2.552$,
and for $N \rightarrow \infty$ tends to $2.481$.
In case of $m=2$, that is, using correlation functions of
the $E_N^{(N-2)}$ type, we have found that
the states $|{\cal S}_n\rangle$ violate inequality (\ref{CHSHg})
for any $N$.
The maximum violation for $N=4$ is $2.418$,
for $N=5$ is $2.424$ and
for $N \rightarrow \infty$ tends to $2.481$.

(d) Nonseparability is robust against the loss of any number of
parties if they can publicly announce the results of their
measurements. For instance,
let us suppose that
$N-2$ observers measure the spin
along the same direction $A$ and publicly announce their results.
Then, if the missing results are $j$ and $k$,
the state shared by the remaining two observers is
\be
|\sigma\rangle_N\!=\!{1 \over \sqrt{2}}
\left( |S_A\!=\!j,S_A\!=\!k\rangle_N\!-\!|S_A\!=\!k,S_A\!=\!j\rangle_N \right).
\ee
Note that $|\sigma\rangle_N$ is formally similar
to the singlet state of two qubits.
However, it belongs to
the $N^2$-dimensional Hilbert
space ${\cal H}_N \otimes {\cal H}_N$
and not to ${\cal H}_2 \otimes {\cal H}_2$,
and therefore does not exhibit rotational symmetry.
For Peres' observables,
the corresponding correlation function is
\be
E_N^{\sigma}(A,B)=(-1)^N \cos^{N-1} \theta_{AB}.
\ee
The states $|\sigma\rangle_N$ violate the
CHSH inequality. For $N=2$ the maximum violation
is $2 \sqrt{2}$, for $N=3$ the maximum violation
is $2.414$, and
for $N \rightarrow \infty$ tends to $2.324$.

A similar situation occurs when any number $p$ of observers (not necessarily two)
measure the same spin component and publicly announce their results.
Then, the state shared by the remaining $N-p$ observers is formally
similar to $|{\cal S}_{N-p}\rangle$, but belongs to
a $N^2$-dimensional Hilbert
space ${\cal H}_N \otimes {\cal H}_N$.
Therefore, in the secret sharing scenario,
if some of the parties get caught by the enemy,
and they are nevertheless able to publicly announce their results, the remaining
parties still share pseudo $|{\cal S}_{N-p}\rangle$ states and
could still use them for secret sharing.

%%%%%%%%%%%%%% Final comments and open problems %%%%%%%%%%%%%%%

Only very recently has it been possible to prepare
optical analogs to the singlet states of {\em two} $N$-level systems
for every $N$ \cite{LHB01} and to test Bell inequalities for two
qutrits \cite{HLB01}.
So far, of the
states $|{\cal S}_N\rangle$,
only the simplest one, the singlet state of two qubits,
has been created in a laboratory.
Preparing these states for $N \ge 3$ is
a formidable physical challenge. The aim of this Letter
has been to point out some
potential applications
of these states
in order to stimulate
the interest in that challenge.

%%%%%%%%%%%%%%%%%%%%% Acknowledgements %%%%%%%%%%%%%%%%%%%%%%

The author thanks N. David Mermin for stimulating conversations,
the Spanish Ministerio de Ciencia y Tecnolog\'{\i}a Grant
No.~BFM2001-3943, the Junta de Andaluc\'{\i}a Grant No.~FQM-239
%and the organizers of the Conference {\em Quantum Information:
%Quantum Entanglement} (Sant Feliu de Gu\'{\i}xols, Spain, 2002)
for support.

%%%%%%%%%%%%%%%%%%%%% Bibliography %%%%%%%%%%%%%%%%%%%%%%%%%%


\begin{thebibliography}{99}
\bibitem{commmurder}
Or in more general, noncriminal, terms:
each of $N$ parties must perform one among $N$ different tasks which
they do not want to perform
themselves. However, each party
volunteers to perform some other party's task.
\bibitem{commown}
$A_i$ has a probability of $1/N$ to have to kill his/her
own desired victim $B_i$. In case $A_i$ must kill $B_i$, since the
fact that $A_i$ wants $B_i$ dead is presumably known by others, the
best thing $A_i$ could do is to participate in another round with
a different set of strangers.
\end{thebibliography}
\end{document}